\documentclass[10pt,conference]{IEEEtran}

\usepackage[utf8]{inputenc}
\usepackage{xspace}
\usepackage{graphicx}
\usepackage{paralist} %
\usepackage{booktabs} %
\usepackage{hyperref}
\usepackage[disable]{todonotes} %
\usepackage{tikz}
\usepackage{listings}
\usepackage{caption}
\usepackage{subcaption}
\usepackage{framed}
\usepackage{mathtools}
\usepackage{multirow}
\usepackage{amssymb}
\usepackage{threeparttable}

\newcommand{\ibox}[1]{%
  \leavevmode\vphantom{#1}%
  \smash{\fboxsep=1pt\fcolorbox{black}{white!50}{#1}}%
}

\definecolor{darkgreen}{rgb}{0,0.5,0}
\definecolor{darkblue}{rgb}{0,0,0.5}
\definecolor{darkred}{rgb}{0.5,0,0}
\usepackage{tikz}
\usetikzlibrary{matrix,arrows,positioning,backgrounds,fit,patterns,shadows} 
\tikzstyle{every picture}+=[remember picture]
\tikzset{
  feature/.style={draw, inner sep=1.5mm, font=\footnotesize\itshape, fill=white, drop shadow},
  opt/.style={fill=white},
}

\graphicspath{{./graphics/}}

\hypersetup {
  urlbordercolor=1 1 1, %
}

\newcommand{\tool}[1]{\textsc{#1}}

\newcommand\feature[1]{{\sc #1}}
\newcommand\code[1]{{\small\sf #1}}

\begin{document}

\title{On the Relation of External and Internal\\Feature Interactions: A Case Study}

\author{\IEEEauthorblockN{Sergiy Kolesnikov}
  \IEEEauthorblockN{Sven Apel}
  \IEEEauthorblockA{University of Passau}
  \and
  \IEEEauthorblockN{Norbert Siegmund}
  \IEEEauthorblockA{Bauhaus-University Weimar}
  \and
  \IEEEauthorblockN{Christian K{\"a}stner}
  \IEEEauthorblockA{Carnegie Mellon University}
}

\maketitle

\begin{abstract}
Detecting feature interactions is imperative for accurately predicting performance of highly-configurable systems.
State-of-the-art performance prediction techniques rely on supervised machine learning for detecting feature interactions, which, in turn, relies on time consuming performance measurements to obtain training data.
By providing information about potentially interacting features, we can reduce the number of required performance measurements and make the overall performance prediction process more time efficient.
We expect that the information about potentially interacting features can be obtained by statically analyzing the source code of a highly-configurable system, which is computationally cheaper than performing multiple performance measurements.
To this end, we conducted a qualitative case study in which we explored the relation between control-flow feature interactions (detected through static program analysis) and performance feature interactions (detected by performance prediction techniques using performance measurements).
We found that a relation exists, which can potentially be exploited to predict performance interactions.
\end{abstract}

\IEEEpeerreviewmaketitle

\section{Introduction}
\label{sec:intro}

A \emph{feature} is an end-user-visible behavior or characteristic of a (software) product that satisfies a stakeholder’s requirement~\cite{KCH+90}.
Features are used to guide structure, reuse, and variation through the development of highly-configurable software systems~\cite{apel2013feature}.
While facilitating the development of highly-configurable software, reducing  development costs, and improving product quality, making features optional introduces new challenges, such as the feature interaction problem~\cite{apel2013exploring}.
A \emph{feature interaction} occurs when the functionality of a feature or its non-functional properties (e.g., performance) are influenced by the presence or absence of one or more other features~\cite{apel2013feature}.
The presence of feature interactions hinders program comprehension and compositional reasoning about the functional and non-functional properties of features.
That is, we cannot reason about the properties of a system configuration (i.e., a valid feature combination) in terms of a straightforward combination of the \emph{individual} influences of the involved features on these properties.
This is because we also have to consider the influences of possible interactions.
A common practical scenario is searching for the best configuration of a system with respect to performance.
To identify this configuration for a given operational environment, we need to know not only the individual influences of the involved features on performance, but also which interactions among these features exist and what influence on performance they have.\looseness=-1

The problem of detecting feature interactions and quantifying their influence on performance has been addressed in the past by employing machine learning~\cite{siegmund2012predicting, GC+13, zhang2015performance}.
For building a training dataset and identifying interactions, these techniques rely on selecting a representative subset from all system configurations (i.e., sampling) and on measuring the performance of each configuration in this sample~(Sec.~\ref{sec:learning}).
The time needed to perform the measurements often makes up a substantial part of the overall time required by machine learning~\cite{SRA13}.
Therefore, reducing the measurement effort---by concentrating on system configurations that potentially have feature interactions---can make these techniques more time efficient and accurate~\cite{siegmund2012predicting}.

The main question that we address in this paper is whether we can efficiently extract information about potentially existing feature interactions, which then can be used in performance prediction, which is, in turn, imperative to guide maintenance and evolution tasks.
In our previous work~\cite{apel2013exploring}, we described two types of interactions:
(1) \emph{external feature interactions}, which can be identified by observing the external behavior of a system, such as performance; and (2) \emph{internal feature interactions}, which can be identified by analyzing or interpreting the source code of a system, for example, using control-flow analysis~\cite{apel2013exploring}.
Our main hypothesis here is that there is a relation between internal and external interactions, and that we can make use of this relation to automatically identify external interactions by identifying internal interactions in a fast and efficient way.
For example, multiple function calls from one feature to another (internal feature interactions) can result in a performance overhead. This performance overhead is present only if both---the caller and the callee features---are present in a configuration (external feature interaction).
This way, the internal interaction is related to its external counterpart.
This relation, if present, would give us hints about the existence of external feature interactions based on the internal ones.
In this work, we follow up on this idea and report on at exploratory case study in which we investigated the control flow among features and its relation to performance feature interactions.
We conjecture that by supplying the performance-prediction procedure with hints about which feature combinations are more likely or less likely to exhibit external feature interactions, the procedure can be made more focused on finding actual interactions.

Taking into account the \emph{exploratory nature} of our study, the \emph{qualitative character} of the expected results, as well as substantial technical challenges, we chose a \emph{case-study approach} (see Sec.~\ref{sec:research-method}) as our research method and two systems---the \tool{mbedTLS} encryption library and the \tool{SQLite} database engine---as non-trivial, real-world subject systems.\footnote{\url{https://tls.mbed.org/}\hspace{1ex}\url{https://www.sqlite.org/}}
\tool{mbedTLS} and \tool{SQLite} are highly-configurable systems used by several large projects, such as \tool{OpenVPN} and \tool{Firefox},\footnote{\url{https://openvpn.net/}\hspace{1ex}\url{https://www.mozilla.org/}} which makes our case study practice-oriented.

In a nutshell, using a recent machine learning technique (Sec.~\ref{sec:learning}), we learned the external (performance) feature interactions among the features of the subject systems.
Furthermore, we manually inspected the code of the systems and confirmed that the learned performance interactions actually exist and that they are actually caused by the interplay of the corresponding features, and not just misinterpreted artefacts of measurement bias or environment noise.

Using a variability-aware control-flow analysis augmented by manual code inspection (Sec.~\ref{sec:typechef}), we identified control-flow interactions among the features of \tool{mbedTLS} and \tool{SQLite}.
That is, we identified the code locations where the features pass the control to one another.

Comparing the set of internal (control-flow) interactions with the set of external (performance) interactions  revealed that those features that interact internally also interact externally (Sec.~\ref{sec:results-relation}), which is in line with our expectation.
Using the identified relation, we were able to substantially shrink the search space of performance feature interactions (Sec.~\ref{sec:discussion}).
Furthermore, we made first steps towards developing an automated predictor for identifying features that are likely to interact externally based on the set of internal interactions, although, with mostly negative results~(Sec.~\ref{sec:fis}).
To the best of our knowledge, this is the first case study that analyzed both the external and the internal feature interactions for the same systems and investigated possible connections between these two types of interactions.

The contributions of this work are the following:
\begin{compactitem}
\item We define a relation between internal and external interactions based on the features these interactions concern, and we discuss the plausibility of this relation.
\item We define a conceptual framework for exploring the relation between internal and external interactions.
\item In a first case study of this kind based on two real-world highly-configurable subject systems, we explore and confirm the relation between internal and external feature interactions.
\item We discuss the implications of our findings for performance prediction of highly-configurable systems.
\end{compactitem}

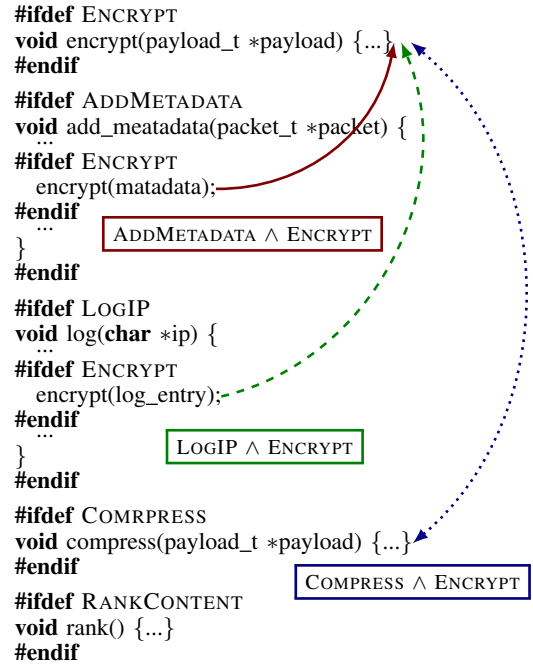
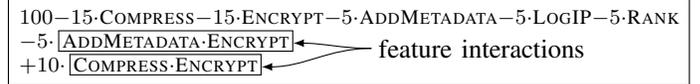
\begin{figure}
  \centering
  \begin{subfigure}{.42\textwidth}
      \begin{lstlisting}
#ifdef @*\feature{Encrypt}*@
void encrypt(payload_t *payload) {...}@*\tikz[baseline=-.5ex]\coordinate(encrypt);*@@*\hspace{.3em}\tikz[baseline=-.5ex]\coordinate(encrypt2);*@@*\hspace{.4em}\tikz[baseline=-.5ex]\coordinate(encrypt3);*@
#endif
@*\vspace{-.6em}*@
#ifdef @*\feature{AddMetadata}*@
void add_meatadata(packet_t *packet) {@*\vspace{-.6em}*@
  ...
#ifdef @*\feature{Encrypt}*@
  encrypt(matadata);@*\tikz[baseline=-.5ex]\coordinate(encrypt_metadata);*@
#endif@*\vspace{-.6em}*@
  ...
}
#endif
@*\vspace{-.6em}*@
#ifdef @*\feature{LogIP}*@
void log(char *ip) {@*\vspace{-.6em}*@
  ...
#ifdef @*\feature{Encrypt}*@
  encrypt(log_entry);@*\tikz[baseline=-.5ex]\coordinate(encrypt_log);*@
#endif@*\vspace{-.6em}*@
  ...
}
#endif
@*\vspace{-.6em}*@
#ifdef @*\feature{Comrpress}*@
void compress(payload_t *payload) {...}@*\tikz[baseline=-.5ex]\coordinate(compress);*@
#endif
@*\vspace{-.6em}*@
#ifdef @*\feature{RankContent}*@
void rank() {...}
#endif
      \end{lstlisting}
    \caption{Control-flow interactions in the audio streaming system.\label{fig:cf-example}}
  \end{subfigure}
  \vspace{1em}
  \begin{tikzpicture}[overlay]
    \path[-latex,line width=1pt,color=darkred] (encrypt_metadata) edge [bend right=40] node [below=0.7em,pos=0.1] {\fboxrule=1pt\fcolorbox{darkred}{white}{\fontsize{8}{8}\selectfont\feature{AddMetadata} $\land$ \feature{Encrypt}}} (encrypt);
    
    \path[-latex,line width=1pt,color=darkgreen] (encrypt_log) edge [bend right=50, dashed] node [below=1.4em,pos=0.1] {\fboxrule=1pt\fcolorbox{darkgreen}{white}{\fontsize{8}{8}\selectfont\feature{LogIP} $\land$ \feature{Encrypt}}} (encrypt2);
    
    \path[latex-latex,line width=1pt,color=darkblue] (compress) edge [bend right=50, dotted] node [below=0.4em,pos=0] {\fboxrule=1pt\fcolorbox{darkblue}{white}{\fontsize{8}{8}\selectfont\feature{Compress} $\land$ \feature{Encrypt}}} (encrypt3);
  \end{tikzpicture}
  \begin{subfigure}{0.5\textwidth}
    \fontsize{7.7}{9.5}\selectfont %
    \begin{framed}
      $100 - 15 \cdot $\feature{Compress}$ - 15 \cdot $\feature{Encrypt}$ - 5 \cdot $\feature{AddMetadata}$ - 5 \cdot $\feature{LogIP}$ - 5 \cdot $\feature{Rank}\\
      $ - 5 \cdot\, $\ibox{\feature{AddMetadata}$ \cdot $\feature{Encrypt}}\tikz[baseline=-.5ex]\coordinate(interaction1);\\
      $ + 10 \cdot\, $\ibox{\feature{Compress}$ \cdot $\feature{Encrypt}}\tikz[baseline=-.5ex]\coordinate(interaction2);
    \end{framed}
    \caption{A performance influence model with performance interactions.\label{fig:model-example}}
    \begin{tikzpicture}[overlay]
      \node (note) at (6.3, 1.32) {\normalsize feature interactions};
      \draw [-latex] (note) to[out=180, in=0] (interaction1);
      \draw [-latex] (note) to[out=180, in=0] (interaction2);
    \end{tikzpicture}
  \end{subfigure}
  \caption{Interactions in the audio streaming system.\label{fig:interaction-example}}
\end{figure}

\section{Internal and External Feature Interactions}

To illustrate how features may interact internally and externally and how these interactions can be related, we use a simple example of an audio streaming system with five optional features:
\feature{Compress} compresses the audio stream;
\feature{Encrypt} encrypts data;
\feature{AddMetadata} adds data about the stream quality, description of the audio content, information about its authors, etc., to the stream;
\feature{LogIP} logs IPs of the users receiving the stream;
\feature{RankContent} ranks the audio content according to its popularity. The performance of the system is measured by the maximum number of users that can simultaneously receive an audio stream without the system becoming overloaded.

\subsection{Control-Flow Interactions (Internal)}
\label{sec:example-control-flow-interactions}
In Figure~\ref{fig:cf-example}, we illustrate an excerpt of the implementation of the audio streaming system.
The code of each feature is delimited using C preprocessor \code{\#ifdef} annotations.
We denote internal interactions among features with arrows.
The boxes on the arrows contain \emph{presence conditions} for the corresponding interactions~\cite{von2015presence}, that is, which features must be enabled (or disabled) for the interaction to take place. For example, if both features \feature{AddMetadata} and \feature{Encrypt} are enabled, then metadata are encrypted along with the audio data.
For this purpose, \feature{AddMetadata} calls the encryption function of feature \feature{Encrypt} (denoted by the solid red arrow).
Consequently, there is a control-flow interaction between these two features.

Likewise, there is a control-flow interaction between features \feature{LogIP} and \feature{Encrypt} (denoted by the dashed green arrow), since the log entries are encrypted if both features are enabled.

Finally, an internal interaction exists between features \feature{Compress} and \feature{Encrypt} (denoted by the dotted blue arrow). This is a data-flow interaction, because both features operate on the same resource (i.e., the audio stream).

\subsection{Performance Interactions (External)}
\label{sec:example-performnce-interactions}

In Figure~\ref{fig:model-example}, we show a \emph{performance influence model}~\cite{siegmund2015performance} of the audio stream system.
For a given system configuration, the model can predict the maximum number of users that can simultaneously receive an audio stream without the system becoming overloaded.\footnote{Here, we assume that the model is 100\,\% accurate.}
To calculate the predicted value, we substitute $1$ for the names of the enabled features and $0$ for the names of all disabled features.
Then, we evaluate the arithmetic expression.
For example, for the configuration with feature \feature{Compress} enabled and the rest of the features disabled the system can reliably serve $100 - 15 \cdot 1 - 15 \cdot 0 + \dots + 10 \cdot 0 = 85$ users.\looseness=-1

The individual terms of the model (i.e., the summands) describe the influence of individual features \emph{as well as} of their interactions on the performance of the system.
The first term ($100$) describes the performance of the base configuration (with all features disabled). The second term ($- 15 \cdot $\feature{Compress}) describes the influence of feature \feature{Compress} on the performance \emph{relative} to the performance of the base system. Thus, the computationally expensive feature \feature{Compress} reduces the base performance by 15.

The terms containing more than one feature (denoted by boxes in Figure~\ref{fig:model-example}) describe the influences of the interactions among the involved features on performance.
For example, enabling both features \feature{AddMetadata} and \feature{Encrypt} makes the system encrypt not only the audio stream, but also the metadata that are added to the stream.
This results in a computational overhead, which reduces the system's performance by 5 users that can be served.

In our example, we assume that encrypting a small string containing an IP address is so fast that this has no measurable effect on the performance of the system.
Therefore, there is no a performance interaction between features \feature{LogIP} and \feature{Encrypt}.
Consequently, there is no a corresponding term in our performance influence model.

The last term in the model describes an interaction between features \feature{Compress} and \feature{Encrypt} with a positive influence of the performance.
Each of the two features individually has a negative influence of $-15$ on the system performance,
but encryption is faster if the data were compressed before.
Therefore, the combined influence of both features on performance is less than the sum of their individual influences: $- 15 - 15 + 10 = -20$ and not $-30$.

Finally, feature \feature{RankContent} as well as all other possible feature combinations have no measurable influence on performance and, therefore, they are not in the performance influence model.

\subsection{Relating Control-Flow and Performance Interactions}
\label{sec:relating-example}

Table~\ref{tab:interactions} summarizes the control-flow and the corresponding performance interactions from our example~(Fig.~\ref{fig:interaction-example}).
The feature combinations (\feature{AddMetadata}, \feature{Encrypt}) and (\feature{Compress}, \feature{Encrypt}) give rise to both control-flow and performance interactions.
Based on our knowledge about the implementation, we can explain the causal relation between the control-flow and performance interactions captured by these feature combinations: The call to the computationally expensive encryption functionality (a control-flow interaction) leads to the performance decrease in the configurations containing the features that implement and use the encryption functionality (i.e., a performance interaction between these features occurs).
Notice that the related control-flow and performance interaction involve exactly the same features, so we can also relate them based on the features they involve.
However, the mere presence of control flow among features does not always indicate the presence of a performance interaction.
For example, the control-flow interaction between features \feature{LogIp} and \feature{Encrypt} has no corresponding performance interaction.
So, it is an open question to what extent a presence of a control-flow interaction can be used as an \emph{indicator} for a potentially existing performance interaction.

\begin{table}
  \centering
  \small
  \caption{A lists of interacting features from Figure~\ref{fig:interaction-example}. It illustrates which of the features interact internally (control-flow interaction), externally (performance interaction), or both.\label{tab:interactions}}
  \begin{tabular}{lcc}
    \toprule
    Interacting Features & Control flow & Performance\\
    \midrule
    \feature{AddMetadata}, \feature{Encrypt} &\checkmark&\checkmark\\ 
    \feature{Compress}, \feature{Encrypt} &\checkmark&\checkmark\\
    \feature{LogIp}, \feature{Encrypt} &\checkmark&--\\
    \bottomrule
  \end{tabular}
\end{table}

\renewcommand*{\thefootnote}{\fnsymbol{footnote}}
Also note that from 26\footnote[1]{10 combinations with 2 features, 10 with 3, 5 with 4, and 1 with 5.}
 feature combinations possible in the audio streaming system only three combinations give rise to feature interactions.
All remaining feature combinations can be ignored by an interaction detection technique, because features in these combinations do not interact.
\renewcommand*{\thefootnote}{\arabic{footnote}}

In what follows, we investigate \emph{to what extent} a relation between control-flow and performance interactions exists in a real-world setting.
Furthermore, we define and evaluate a predictor that uses control-flow interactions to predict potential performance interactions.
With such a predictor in place, we could make interaction detection more efficient and accurate, which would be a valuable contribution to research fields, such as optimization of non-functional properties, combinatorial testing, and sampling techniques.

\section{Conceptual Framework}

Next, we describe the methods and tools used in our study and how we combined them in a conceptual framework to study relations among internal and external interactions.

\subsection{Research Method}
\label{sec:research-method}

Our study is explorative in nature, because the way we study the relation between control-flow and performance interactions is novel and requires careful validation in itself.
Our study involves bleeding-edge techniques for detecting control-flow and performance interactions that are technically challenging and, therefore, cannot be easily applied to a large number of non-trivial real-world systems.
By focusing on two systems, we aim at increasing the internal validity of the study, because, this way, we can better identify and control confounding effects that may vary substantially from one subject system to another (e.g., architecture, size of features).
Ultimately, we want to obtain deep insights in the nature of the relation between control-flow and performance feature interactions and not only report on shallow statistics.
Taking these characteristics of our study into account, the literature on conducting empirical research in software engineering suggests a \emph{case study} as a research method: Shull et al.\ describe case studies as ``initial investigations of some phenomena''~\cite{Shull2007}, Yin extends the description by stating that a case study is ``an empirical inquiry that investigates a contemporary phenomenon within its real-life context, especially when the boundaries between phenomenon and context are not clearly
evident''~\cite{yin2003case}, and Flyvbjerg adds that ``case studies offer in-depth understanding of how and why certain phenomena occur, and can reveal the mechanisms by which cause–effect relationships occur''~\cite{flyvbjerg2006five}.

\subsection{Identifying Control-Flow Interactions}
\label{sec:typechef}
To identify control-flow interactions, we use a variability-aware call-graph analysis~\cite{ferreira2015characterizing} %
implemented in \tool{TypeChef}\footnote{\url{http://fosd.net/TypeChef/}} that identifies function calls among features implemented with preprocessor annotations (Fig.~\ref{fig:cf-example}).
The central idea of a variability-aware analysis is to achieve efficiency by analyzing code parts that are shared by multiple system configurations only once.
This is achieved by analyzing the source code of the system that still contains variability (e.g., the code with preprocessor annotations in Figure~\ref{fig:cf-example}), as opposed to analyzing the source code of individual configurations, which may be exponentially many in the number of features. A variability-aware call-graph analysis provides an efficient way to identify function calls among features of a highly-configurable system and makes the detection of internal interactions feasible.

The underlying data structure for the analysis is \emph{the variable abstract syntax tree}.
Similar to an abstract syntax tree (AST), a variable AST provides an abstraction of the source code that can be efficiently analyzed, but it also provides information on which part of the code belongs to which features (in the form of presence conditions).
Using this information, a call-graph analysis can identify, for each function call, which feature is the caller and which feature is the callee.
Furthermore, the analysis can identify a presence condition for each call, that is, which features must be enabled (or disabled) for the call to take place at runtime.
For example, in Figure~\ref{fig:cf-example}, the call from feature \feature{AddMetadata} to feature \feature{Encrypt} (solid red arrow) occurs only if both features \feature{AddMetadata} and \feature{Encrypt} are enabled (denoted by the presence condition in the box under the arrow).
Due to the static nature of the technique, the collected information about the calls may be an overapproximation, but this is a problem with any static analysis approach.
The current implementation of the analysis also uses pointer analysis to increase the accuracy of the call graph~\cite{ferreira2015characterizing}.\looseness=-1

\subsection{Identifying Performance Interactions}
\label{sec:learning}
For detecting performance feature interactions, we learn performance influence models (Fig.~\ref{fig:model-example}).
As discussed in Section~\ref{sec:example-performnce-interactions}, a performance influence model captures the influences of individual features and their interactions on performance of a configurable system.
We learn performance influence models using the tool \tool{SPL~Conqueror},\footnote{\url{http://fosd.net/SPLConqueror/}} which implements a state-of-the-art machine learning algorithm based on multivariate regression and forward feature selection~\cite{siegmund2015performance}.
The algorithm takes as input a sample of system configurations and corresponding performance measurements.
The accuracy of the learned performance influence model depends, among other factors, on how representative the sampled configurations are for the entire configuration space.
To get a performance influence model of the highest possible accuracy, and, consequently, to detect feature interactions as precise as possible (i.e., to obtain the ground truth), we measured not a sample but all configurations of the subject system and used these measurements as the algorithm input.
The performance measurements were done using a standard benchmark.

To build an influence model, \tool{SPL~Conqueror} starts with calculating a set of features and their combinations that can be included in the model to reduce the model's prediction error.
For example, \feature{Compress}$\,\cdot\,$\feature{Encrypt} in Figure~\ref{fig:model-example} is a feature combination that has been eventually included in the model during the learning process.
The algorithm iterates over the set of features and their combinations and selects one element of the set that explains variations in the performance of the system best; that is, the element yielding the model's lowest prediction error, when incorporated into the model.
The selection of candidates %
continues until either a predefined accuracy is reached or all features and feature combinations that could reduce the prediction error of the model have been considered.
For a more in-depth description of the algorithm, we refer the reader to
previous work~\cite{siegmund2015performance}.

\subsection{Relating Control-Flow and Performance Interactions}
\label{sec:relation-method}

After we have identified the internal (control-flow) interactions, the question is what we can learn from them regarding external (performance) interactions.
To answer this question, we relate the control-flow interactions and performance interactions based on the features involved in them, as we explained it in our example in Section~\ref{sec:relating-example}.
The goal is to find out if the features involved in performance interactions also occur in one or more internal interactions and vice versa.
This is a feasibility check to see if the interactions can be related based on the features' occurrence at all.
That is, if we find no interactions that can be related in this way, this would mean that it is impossible to define any relation between interactions based on the corresponding feature occurrences in these interactions.

We define a performance interaction $i_p$ and a control-flow interaction $i_c$ as related if $\mathit{features}(i_p) \subseteq \mathit{features}(i_c)$ or if $\mathit{features}(i_p) \supseteq \mathit{features}(i_c)$, where $\mathit{features}(i)$ is the set of features that contribute to the interaction $i$.

Furthermore, for each related pair of interactions, we determine how similar the interactions are (i.e., if they contain exactly the same features or if they also contain features that are present only in one of them).
The similarity of the related interactions can be interpreted as the strength of their relation: the higher the similarity, the higher the strength of the relation.
We calculate the similarity of interactions using the Jaccard index $J$~\cite{jaccard1912distribution}:\vspace{-1ex}\\
\phantom{blablablablablabla}$J(i_p, i_c) = \frac{\mathit{features}(i_p) \cap \mathit{features}(i_c)}{\mathit{features}(i_p) \cup \mathit{features}(i_c)}$\vspace{1ex}\\
where $\mathit{features}(i)$ is the set of features involved in the interaction $i$.
The Jaccard index equals 1 if both interactions involve exactly the same features and is less than 1 otherwise.

\subsection{Predicting Performance Interactions}
\label{sec:fis}

If we find a relation between control-flow and performance feature interactions as defined in Section~\ref{sec:relation-method}, the question is whether we can use this relation to predict performance feature interactions.

One method is to build on our argumentation in Section~\ref{sec:relating-example} and to assume that every control-flow interaction corresponds to an existing performance interactions.
Of course, we already know that there may be control-flow interactions without corresponding performance interactions.
Nevertheless, it is an open question \emph{how} accurate this simple method can be if applied to a real-world system.

We can also use a more advanced method based on reoccurring feature combinations in control-flow interactions: We argue that, if a set of features occurs in multiple control-flow feature interactions, then this set of features is also likely to give rise to one or more external interactions.
The rationale behind this argument is that, if a set of feature is involved in many control-flow feature interactions, then chances are high that it is also involved in performance interactions, because the accumulated influence of the control-flow interactions on performance have a measurable effect.

The method that we use to identify such frequent feature sets is \emph{frequent item set mining}~\cite{borgelt2012frequent}, which has been successfully used as a general pattern mining method~\cite{maqbool2007hierarchical, qiao2013analyzing}.
In terms of frequent item set mining, we refer to a feature as an \emph{item}.
The set of all items (all features) is the \emph{item base} $B$.
A subset of the item base $I \subseteq B$ is an \emph{item set} that corresponds to a feature combination.
An item set (i.e., a feature combination) that denotes an internal interaction in a system is a \emph{transaction} $t \in T$, where $T$ is a set of transactions.
Based on these definitions, we define the \emph{support} (a.k.a. absolute frequency) $s$ of an item set:
$s = \left|\{t : t \in T \land I \subseteq t\}\right|$.
The support value and a threshold $E \in [0,\infty)$ is used to decide which of the item sets are considered frequent:
All item sets with the support value $s \geq E$ are \emph{frequent} item sets.
Based on our hypothesis, frequent item sets predict external feature interactions.
We use an implementation of the Apriori algorithm %
from the \tool{Orange} library\footnote{\url{http://orange.biolab.si/}} to calculate the support value.

\section{Case Study}

We address the following research questions in our study:

\begin{compactitem}
  \item \textbf{RQ1:} Do control-flow feature interactions and performance feature interactions relate (in terms of the definition of Section~\ref{sec:relation-method})?
  \item \textbf{RQ2:} If the relation identified in RQ1 exists, what is the predictive power (i.e., precision and recall) of the predictors that we propose in Section~\ref{sec:fis}?
\end{compactitem}

To answer these research questions, we search for control-flow and performance interactions in the subject systems, study their properties, identify relations between them, and evaluate predictors for performance interactions based on these relations.

\subsection{Subject Systems}
\label{sec:subject-system}
The case study was conducted using two real-world highly-configurable software systems: the \tool{mbedTLS} library implementing the transport security network protocol TLS/SSL and a SQL database engine \tool{SQLite}.
The initial use case for the systems was embedded domain, but now they are  also used in non-embedded projects, such as \tool{OpenVPN} and \tool{Firefox}.

Similar to a large number of other real-world highly configurable systems,
the subject systems are written in C using C-preprocessor directives to implement compile-time variable features.
\tool{mbedTLS} comprises 50\,K and \tool{SQLite} 195\,K lines of code. Both systems have a highly modular architecture, which is thoroughly documented along with the corresponding preprocessor macro names allowing relatively easy matching of code to the corresponding modules and submodules.

\paragraph{\tool{mbedTLS} Features and Feature Model}
At the top level, \tool{mbedTLS} consists of modules, such as \textit{Cipher}, \textit{Public Key}, \textit{Hashing}.
Each module implements the corresponding algorithms and protocols. For example, the \emph{Cipher} module includes submodules that implement cipher algorithms, such as AES, DES, and ARC4.
Submodules implement the features of the system.
The cipher-algorithm features can be combined with other features, such as hash algorithms and public-key implementations, to provide an encryption protocol.
We used the original documentation and manual code inspection to construct a feature model for \tool{mbedTLS} version 2.2.1, comprising 97 features and 1921 configurations.

\paragraph{\tool{SQLite} Features and Feature Model}
\tool{SQLite} consists of a \textit{Core} providing a C-language interface and being responsible for executing compiled SQL code, an \textit{SQL Compiler}, and a \textit{Backend} providing the low-level implementation of the database.
A user can configure the operation of these modules by enabling or disabling their features through compile-time options. For example, \textit{Core} can be configured to operate safely in a multithreaded environment by enabling the \feature{SQLITE\_THREADSAFE} feature.
We studied the documentation and the source code of version 3.16.2 to construct a feature model comprising 12 features and 1533 configurations.

\paragraph{\tool{mbedTLS} Performance Measurements}
The primary application of \tool{mbedTLS} is the encryption of data transmitted over a TCP/IP network.
Ensuring fast and secure data transfer is commonly considered an important property of communication networks, such as the Internet.
So, the time required to encrypt data and transfer them over the network is an important non-functional property of \tool{mbedTLS}.
Measuring the time required by encryption alone is not representative, because different configurations may produce different amounts of payload (e.g., due to data compression and different amounts of generated metadata) influencing the transmission time.
Therefore, we defined the performance measure for a configuration of \tool{mbedTLS} as the amount of time (in seconds) required to encrypt and successfully transmit a fixed amount of input data.

To detect performance feature interactions reliably based on performance benchmarks, it must be ensured that every feature included in a configuration is invoked during the benchmark of this configuration.
Otherwise, the influence of features and their interactions on performance cannot be deduced from the benchmark results.
The original automated test framework of \tool{mbedTLS} includes tests that check the library's functionality in a client-server environment and is suitable to serve as a typical benchmark suite.
During the tests, the functionality of every feature in the configuration is tested, that is, every feature is actually invoked.

We used 2\,GB of random data as input to ensure that the fastest configuration requires, at least, five seconds for transmission and to mitigate the influence of warm-up effects on the result.
We repeated the benchmark 30 times to further reduce the influence of measurement bias.
To exclude the influence of network latencies, we ran the benchmark locally using the local network interface.

\paragraph{\tool{SQLite} Performance Measurements}
The developers of \tool{SQLite} provide a performance benchmark that measures time required by the database to execute a set of queries.\footnote{http://sqlite.org/speed.html}
The original benchmark is not compatible with the latest version of the system that we use, so we used it as guidance to create a new compatible benchmark.
While constructing the benchmark we made sure that the features of \tool{SQLite} are actually invoked during the benchmarking process.
Our benchmark measures the execution time in seconds.
To reduce the influence of warm-up effects and measurement bias, the benchmark runs, at least, 25 seconds and every run is repeated 30 times.\looseness=-1

The benchmarks for both systems were conducted on an Intel i5-4590, 16\,GB RAM, 256\,GB SSD, Ubuntu\,16.04.

\subsection{Results}
\label{sec:results}

\subsubsection{Performance Interactions}
\label{sec:results-performance}
We used \tool{SPL~Conqueror} and the performance benchmark results (cf. Sec.~\ref{sec:subject-system}) as input data to identify performance interactions in \tool{mbedTLS} and \tool{SQLite}, as described in Section~\ref{sec:learning}.
Table~\ref{tab:result-perf-int} lists for both systems the performance interactions and their influences on performance of the systems in seconds.
The \emph{negative values} in the influence column denote \emph{positive influences} of the corresponding interactions on performance.
That is, they denote how much less time a configuration that includes them would need to execute the benchmark.

The mean standard deviation for the performance measurements of \tool{mbedTLS} is 0.42\,s.
Therefore, we classified all interactions with the absolute influences less than this value as noise and discarded them.
From the remaining 16 interactions, 11 are interactions between two features; and five are interactions among three features.
The mean standard deviation for the performance measurements of \tool{SQLite} is 0.09\,s.
The influences of the three identified interactions for the system are much higher and, therefore, are unlikely to be noise. Two of the interactions are interactions between two features and one is an interaction between three features.

\begin{table}
  \begin{threeparttable}
    \centering
    \footnotesize
    \caption{Performance interactions, their influences on performance of the systems in seconds, the number of the control-flow interactions related to them, and the mean value of the corresponding Jaccard indexes. (') marks the relations for manually added control-flow interactions.\label{tab:result-perf-int}}
    \begin{tabular}{@{\hspace{0\tabcolsep}} l @{\hspace{.5\tabcolsep}} r @{\hspace{1\tabcolsep}} r @{\hspace{1\tabcolsep}} l @{\hspace{-.5ex}} r @{\hspace{1\tabcolsep}} r @{\hspace{0.2\tabcolsep}}}
      \toprule
      & ID & Influence & Performance Interaction & \multicolumn{1}{l}{Rela-} & Jaccard \\
      & & \multicolumn{1}{l}{(sec)} & \multicolumn{1}{l}{(features involved)} & \multicolumn{1}{l}{tions} & (mean) \\
      \midrule %
      \multirow{20}{*}{\rotatebox[origin=c]{90}{\textbf{\tool{mbedTLS}}}} & 1  &10.73 & {\scriptsize\feature{CIPHER\_MODE\_CBC}, \feature{SHA256\_C}} & 1' & 1.00 \\
      & 2  &-9.71 & {\scriptsize\feature{AES\_C}, \feature{AESNI\_C}} & 10\phantom{'} & 0.53 \\
      & 3  & 8.53 & {\scriptsize\feature{AESNI\_C}, \feature{SSL\_CBC\_RECORD\_SPLITTING}} & 2\phantom{'} & 0.38 \\
      & 4  & 6.93 & {\scriptsize\feature{CIPHER\_MODE\_STREAM}, \feature{AESNI\_C}} & 1' & 1.00 \\
      & 5  & 6.08 & {\scriptsize\feature{SHA256\_C}, \feature{CIPHER\_MODE\_STREAM}} & 1' & 1.00 \\
      & 6  & 5.75 & {\scriptsize\feature{AES\_C}, \feature{AESNI\_C}, \feature{GCM\_C}} & 4\phantom{'} & 0.53 \\
      & 7  & 3.49 & {\scriptsize\feature{CIPHER\_MODE\_CBC}, \feature{SHA256\_C},} & 1' & 1.00 \\
      & &  & {\scriptsize\feature{SHA256\_SMALLER}} &  & \\
      & 8  & 3.45 & {\scriptsize\feature{SHA256\_C}, \feature{CIPHER\_MODE\_STREAM},} & 1' & 1.00 \\
      & &  & {\scriptsize\feature{SHA256\_SMALLER}} &  & \\
      & 9  & 3.44 & {\scriptsize\feature{SHA256\_C}, \feature{AESNI\_C},} & 1' & 1.00 \\
      & &  & {\scriptsize\feature{CIPHER\_MODE\_STREAM}} &  & \\
      & 10 & 3.14 & {\scriptsize\feature{CIPHER\_MODE\_CBC}, \feature{RIPEMD160\_C}} & 1' & 1.00 \\
      & 11 &-2.97 & {\scriptsize\feature{AES\_C}, \feature{GCM\_C}} & 13\phantom{'} & 0.40 \\
      & 12 &-2.84 & {\scriptsize\feature{CIPHER\_MODE\_STREAM}, \feature{ MD5\_C}} & 1' & 1.00 \\
      & 13 & 1.93 & {\scriptsize\feature{AESNI\_C}, \feature{CAMELLIA\_C}} & 4\phantom{'} & 0.35 \\
      & 14 & 1.68 & {\scriptsize\feature{CIPHER\_MODE\_CBC}, \feature{SHA1\_C}} & 1' & 1.00 \\
      & 15 & 1.60 & {\scriptsize\feature{CIPHER\_MODE\_STREAM}, \feature{AESNI\_C},} & 1' & 1.00 \\
      & &  & {\scriptsize\feature{MD5\_C}} &  & \\
      & 16 & 1.51 & {\scriptsize\feature{RIPEMD160\_C}, \feature{CIPHER\_MODE\_STREAM}} & 1' & 1.00 \\
      \midrule
      \multirow{4}{*}{\rotatebox[origin=c]{90}{\textbf{\tool{SQLite}}}} & 1 & 1.50 & {\scriptsize\feature{DEFAULT\_MEMSTATUS}, \feature{THREADSAFE}} & 1' & 1.00 \\
      & 2  &1.47& {\scriptsize\feature{MEMDEBUG}, \feature{THREADSAFE}} & 16\phantom{'} & 0.45 \\
      & 3  &1.41& {\scriptsize\feature{DEFAULT\_MEMSTATUS}, \feature{MEMDEBUG},} & 1' & 1.00\\
      & &  & {\scriptsize\feature{THREADSAFE}} &  & \\
      \bottomrule
    \end{tabular}
    \begin{tablenotes}
      \footnotesize
      \item To relate the influences to configuration run times, note that the fastest \tool{mbedTLS} configuration completed its benchmark in 6.7 seconds and the fastest \tool{SQLite} configuration completed its benchmark in 26.7 seconds.
    \end{tablenotes}
  \end{threeparttable}
\end{table}

\paragraph{\tool{mbedTLS}}
All identified interactions in \tool{mbedTLS} are among features implementing different ciphers, block cipher modes of operation (simply ``modes''), and cryptographic hash functions.
This is plausible, because these three types of algorithms work tightly together to implement an encryption protocol.
Ciphers (e.g., AES) are used to encrypt data, modes (e.g., CBC) are used in combination with block ciphers to encrypt amounts of data larger than a block (i.e., a fixed amount of data a block cipher can operate on; 128\,bit for AES), and cryptographic hash functions (e.g., SHA) are used with modes to implement authentication and to ensure data integrity.\looseness=-1

To confirm that the identified performance interactions actually result from the interplay of the corresponding features, we manually inspected the source code of \tool{mbedTLS}. 
Next, we present the results of this code inspection.

Interaction 1 in Table~\ref{tab:result-perf-int} arises between a mode (CBC) and a hash function (SHA256).
CBC uses hashing extensively to calculate keyed-hash message authentication code (HMAC).
SHA256 is computationally more expensive than, for example, MD5; therefore, this combination with the mode has a negative influence of $10.73$ seconds on performance.
Interactions 5, 7, 8, 10, 12, 13, 14, and 16 have a similar cause and explanation.
In addition to a mode and a hash function, interactions 7 and 8 also include the feature \feature{SHA256\_SMALLER}, which denotes that an implementation of SHA256 with smaller binary footprint was used.
However, this implementation also has a lower performance, which leads to the negative influence of this interaction on performance.
Interaction 12 has a positive influence on performance of using a mode (stream mode, in this case) with a less computationally complex (but also less secure) MD5 hash function.
In interaction 13, the AES cipher is used as a hash function in combination with the Camelllia cipher.\looseness=-1

Interaction 2 arises from the usage of the AES cipher for encryption in combination with a native implementation of the AES algorithm in assembler (AESNI).
The native implementation makes encryption faster, so this interaction has a positive influence of $9.71$ seconds on performance.

Interaction 3 arises from the usage of the AES cipher for encryption in combination with an implementation of the CBC mode that includes a record splitting algorithm. This algorithm is a countermeasure against the BEAST attack on the SSL algorithm.
The way record splitting is implemented increases the number of packets to be transmitted (compared to the number of packets without this countermeasure).
The increased number of packets results, in turn, in a negative influence on performance.

Interactions 4, 6, 9, 11, and 15 arise from the influence of further  combinations of ciphers, modes, and hash functions on performance, similar to the first interaction.

\paragraph{\tool{SQLite}}
All performance interactions in \tool{SQLite} include the feature \feature{THREADSAFE}.
This is plausible, because \feature{THREADSAFE} is a crosscutting feature that adds the mutex and thread-safety logic to all unsafe regions in the code.
This additional thread-safety code imposes a runtime overhead and makes the benchmarks for the configurations containing it run longer.
We inspected the code of \tool{SQLite} and confirmed that both features \feature{DEFAULT\_MEMSTATUS} and \feature{MEMDEBUG} retrieve a mutex (i.e., use \feature{THREADSAFE} feature) at a certain stage of operation that results in interaction among \feature{THREADSAFE} and these features.

\vspace{-1.5ex}\begin{framed}
\paragraph{Summary}
Overall, we identified 16 performance interactions in \tool{mbedTLS}. 
11 of them occur between 2 features and 5 among 3 features.
In \tool{SQLite}, we identified 3 performance interactions. 2 interactions between 2 features and 1 among 3 features.
Using domain knowledge and manual inspection of the source code, we identified the cause of all interactions and thereby confirmed that they actually exist in the systems and are caused by the interplay of the corresponding features.
\end{framed}

\subsubsection{Control-Flow Interactions}
\label{sec:results-control-flow}

We used the variability-aware call-graph analysis implemented in \tool{TypeChef} (Sec.~\ref{sec:typechef}) to detect control-flow interactions in \tool{mbedTLS} and \tool{SQLite}.

\paragraph{\tool{mbedTLS}} From 761\,992 function calls in the system, we detected 575\,560 control-flow feature interactions.
This number of interactions includes duplicate interactions that appear if the corresponding function call between features occurs in multiple locations in the code. The number of unique control-flow interactions is 73.

\begin{figure}
  \begin{subfigure}[t]{.24\textwidth}
    \centering
    \includegraphics[width=\textwidth]{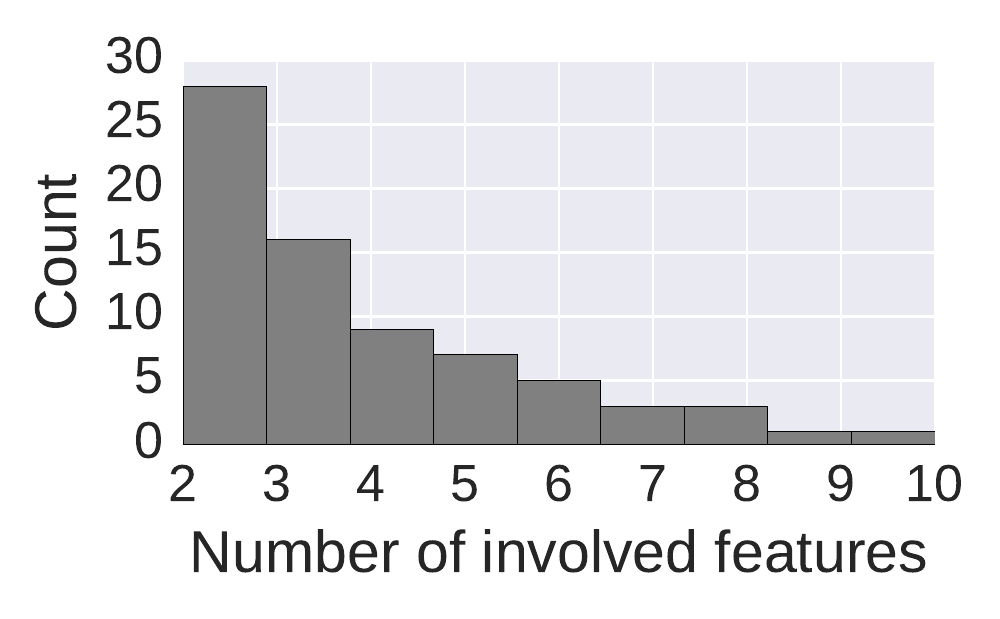}
    \caption{Unique interactions only\label{fig:ci-distribution-unique}}
  \end{subfigure}%
  \begin{subfigure}[t]{.24\textwidth}
    \centering
    \includegraphics[width=\textwidth]{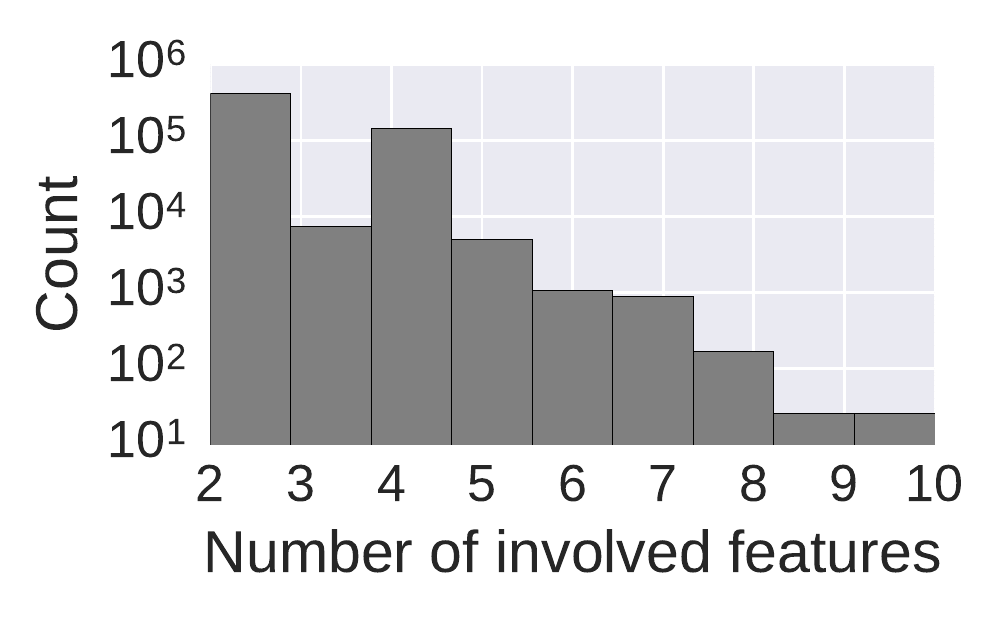}
    \caption{All interactions\label{fig:ci-distribution-all}}
  \end{subfigure}
  \caption{{\small\tool{mbedTLS}}: counts of features in control-flow interactions.\label{fig:ci-distribution}}
\end{figure}

Notably, among the unique control-flow interactions, there are interactions with up to 10 features, but most unique interactions involve only two features (Fig.~\ref{fig:ci-distribution-unique}).
If we also consider the duplicates (Fig.~\ref{fig:ci-distribution-all}), the overall picture stays largely the same: Only the number of interactions involving four features becomes larger than the number of those involving three features.

While manually exploring the source code of \tool{mbedTLS}, we found that cipher, mode, and hash algorithms call each other indirectly, using function pointers.
This indirection was introduced by the designers of the library to decouple the algorithms and to make their concrete implementations interchangeable. \tool{TypeChef} would need to be extended with a variability-aware, inter-procedural data-flow analysis to identify which features interact using indirect function calls.
Being aware of this technical limitation of \tool{TypeChef}, we added 11 indirect control-flow interactions that we collected while manually exploring the code to the set of interactions. Therefore, the total number of the identified unique control-flow interactions is 84 (73 were found using \tool{TypeChef} and 11 manually).
It would be infeasible to find manually all instances of indirect control-flow interactions, so their exact number (including duplicates) is unknown.
We discuss the corresponding threats to validity in Section~\ref{sec:threats}.

\begin{figure}
  \begin{subfigure}[t]{.24\textwidth}
    \centering
    \includegraphics[width=\textwidth]{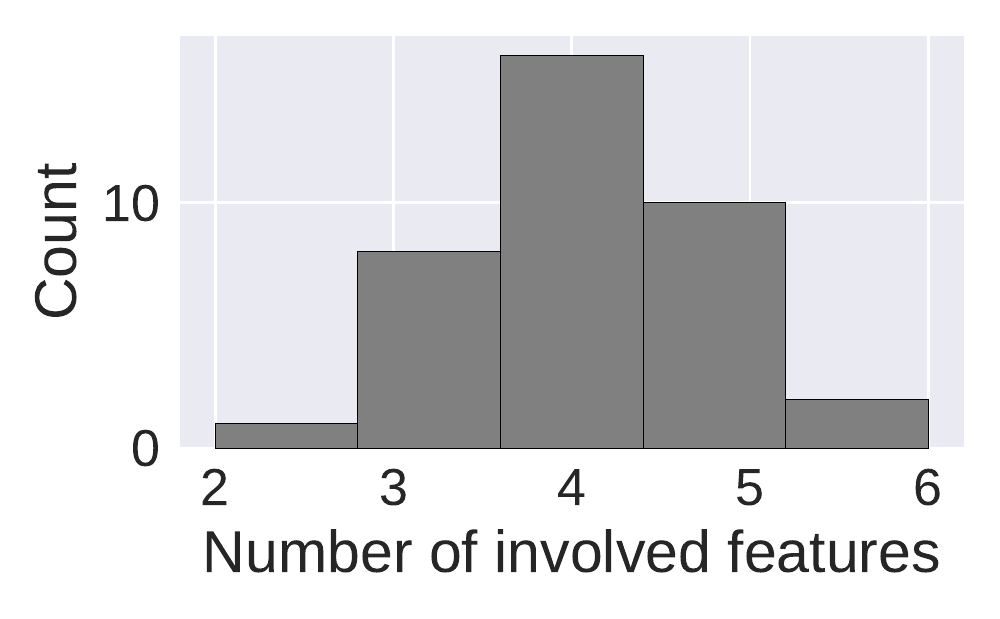}
    \caption{Unique interactions only\label{fig:ci-distribution-unique-sqlite}}
  \end{subfigure}%
  \begin{subfigure}[t]{.24\textwidth}
    \centering
    \includegraphics[width=\textwidth]{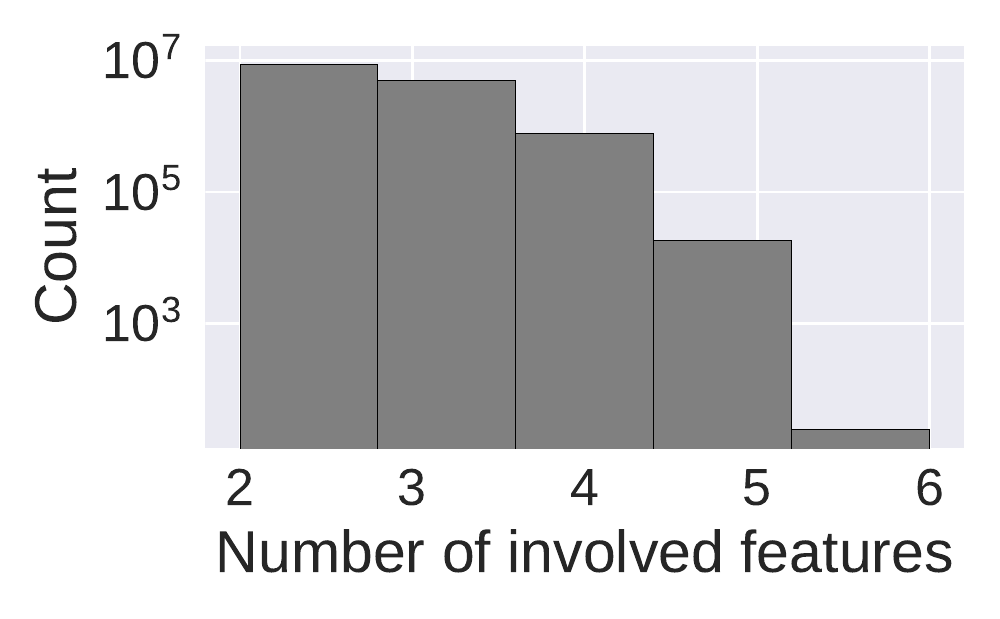}
    \caption{All interactions\label{fig:ci-distribution-all-sqlite}}
  \end{subfigure}
  \caption{{\small\tool{SQLite}}: counts of features in control-flow interactions.\label{fig:ci-distribution-sqlite}}
\end{figure}

\paragraph{\tool{SQLite}} From over 14\,587\,337 function calls in the system, we detected 14\,587\,335 control-flow feature interactions. That is, all but two function calls involved more than one feature. The number of unique control-flow interactions is 37.\looseness=-1

In contrast to \tool{mbedTLS}, most unique interactions involve 4 features, and there are interactions with up to 6 features (Fig.~\ref{fig:ci-distribution-unique-sqlite}). Although, if we also consider duplicates (Fig.~\ref{fig:ci-distribution-all-sqlite}), the picture becomes similar to that in \tool{mbedTLS}: Interactions among 2 features prevail and the count of interactions decreases with the increasing number of involved features.

While manually inspecting the code of \tool{SQLite}, we found that the option \feature{SQLITE\_DEFAULT\_MEMSTATUS} (which is used by \tool{TypeChef} to identify the code belonging to the feature \feature{DEFAULT\_MEMSTATUS}) is used to set a Boolean variable at compile-time.
This variable is then used at runtime to check if feature \tool{DEFAULT\_MEMSTATUS} is enabled or disabled.
This way, the feature can be enabled or disabled at runtime.
Again, \tool{TypeChef} would need a data-flow analysis to trace the connection the preprocessor macro to the corresponding Boolean variable to detect control-flow interactions in which feature \feature{DEFAULT\_MEMSTATUS} is involved.
By further exploring the code, we identified two control-flow interactions of this kind and added them to the set of automatically detected interactions. Therefore, the total number of the identified unique control-flow interactions is 39.

\vspace{-1.5ex}\begin{framed}
\paragraph{Summary}
Overall, we identified 575\,571 control-flow interactions in \tool{mbedTLS} among which 84 were unique. Some interactions involve up to 10 features, but most interactions are between 2 features.
For \tool{SQLite}, we identified 14\,587\,335 control-flow interactions, with 39 unique.
Due to technical limitations of \tool{TypeChef}, indirect control-flow interactions in \tool{mbedTLS} and interactions induced by runtime variability in \tool{SQLite} could not be detected by \tool{TypeChef}.
We manually inspected the source code to collect these interactions.
\end{framed}

\subsubsection{Relating Interactions}
\label{sec:results-relating}

\paragraph{Performance Interactions $\rightarrow$ Control-Flow Interactions.} Using the relation definition $\mathit{features}(i_p) \subseteq \mathit{features}(i_c)$ (Section~\ref{sec:relation-method}), for each performance interaction, we identified all unique related control-flow interactions (i.e., all control-flow interactions involving exactly the same features as the performance interaction).
Furthermore, for each pair of related interactions, we calculated the Jaccard index (Section~\ref{sec:relation-method}), which denotes how similar the interactions are (the index equals 1 if both interactions involve exactly the same features and is less than 1 otherwise).

Table~\ref{tab:result-perf-int} summarizes the results. For each performance interaction, it shows the number of the related control-flow interactions and the mean of all Jaccard indexes calculated for these relations.
The apostrophe (') denotes the performance interactions that are related to the manually identified indirect control-flow interactions for which we were not able to establish the exact number of occurrences (cf.~Section~\ref{sec:results-control-flow}).
The numbers show that there is a relation between every performance interaction and, at least, one control-flow interaction.
The Jaccard indexes show further that the related control-flow interactions, which were automatically detected by \tool{TypeChef}, involve, on average, twice as many or even more features than there are in the corresponding performance interactions.\looseness=-1

\paragraph{Control-Flow Interactions $\rightarrow$ Performance Interactions.}
\label{sec:results-relation}

Using the relation definition $\mathit{features}(i_p) \supseteq \mathit{features}(i_c)$ (Section~\ref{sec:relation-method}), for each control-flow interaction, we identified all related performance interactions (i.e., all performance interactions involving exactly the same features as the control-flow interaction).

Table~\ref{tab:internal_in_external} summarizes the results.
For \tool{mbedTLS}, among the 84 unique control-flow interactions, we found 4 interactions that have one or more related performance interactions.
For \tool{SQLite}, among the 39 unique control-flow interactions, we found 2 interactions that have one or more related performance interactions.
The Jaccard indexes show that the related performance interactions that were automatically detected by \tool{TypeChef} involve mostly the same features as the corresponding control-flow interactions.
The manually added control-flow interactions match exactly the related performance interactions.

\vspace{-1.5ex}\begin{framed}
\paragraph{Summary} We found a relation between every of the 16 identified performance interactions and one or more control-flow interactions.
The Jaccard indexes for the related interactions show that the interactions do not generally contain exactly the same features and that the related control-flow interactions involve, on average, twice as many features than there are in the corresponding performance interactions.
\end{framed}

\begin{table}
  \centering
  \footnotesize
  \caption{Control-flow interactions, the number of the related performance interactions, and the mean value of the corresponding Jaccard indexes. Control-flow interactions without related performance interactions are not listed. (') marks the relations for manually added control-flow interactions.\label{tab:internal_in_external}}
  \begin{tabular}{@{\hspace{.5\tabcolsep}} l @{\hspace{1\tabcolsep}} rl @{\hspace{0\tabcolsep}} rr}
    \toprule
    & ID & Control-Flow Interaction & \multicolumn{1}{l}{Rela-} & Jaccard \\
    & & (features involved) & \multicolumn{1}{l}{tions} & (mean) \\
    \midrule %
    \multirow{18}{*}{\rotatebox[origin=c]{90}{\textbf{\tool{mbedTLS}}}} & 1 & {\scriptsize\feature{AES\_C}, \feature{AESNI\_C}} & 2\phantom{'} & 0.83 \\
    & 2 & {\scriptsize\feature{GCM\_C}, \feature{AESNI\_C}} & 1\phantom{'} & 0.67 \\
    & 3 & {\scriptsize\feature{GCM\_C}, \feature{AES\_C}} & 2\phantom{'} & 0.83\\
    & 4 & {\scriptsize\feature{GCM\_C}, \feature{AES\_C}, \feature{AESNI\_C}} & 1\phantom{'} & 1.00 \\
    & 5 & {\scriptsize\feature{CIPHER\_MODE\_CBC}, \feature{SHA256\_C}} & 1' & 1.00 \\
    & 6 & {\scriptsize\feature{CIPHER\_MODE\_STREAM}, \feature{AESNI\_C}} & 1' & 1.00 \\
    & 7 & {\scriptsize\feature{SHA256\_C}, \feature{CIPHER\_MODE\_STREAM}} & 1' & 1.00 \\
    & 8 & {\scriptsize\feature{CIPHER\_MODE\_CBC}, \feature{SHA256\_C},} & 1' & 1.00 \\
    & & {\scriptsize \feature{SHA256\_SMALLER}} & & \\
    & 9 & {\scriptsize\feature{SHA256\_C}, \feature{CIPHER\_MODE\_STREAM},} & 1' & 1.00 \\
    & & {\scriptsize\feature{SHA256\_SMALLER}} & & \\
    & 10 &{\scriptsize \feature{SHA256\_C}, \feature{AESNI\_C},} & 1' & 1.00 \\
    & & {\scriptsize\feature{CIPHER\_MODE\_STREAM}} & & \\
    & 11 &{\scriptsize \feature{CIPHER\_MODE\_CBC}, \feature{RIPEMD160\_C}} & 1' & 1.00 \\
    & 12 &{\scriptsize \feature{CIPHER\_MODE\_STREAM}, \feature{ MD5\_C}} & 1' & 1.00 \\
    & 13 &{\scriptsize \feature{CIPHER\_MODE\_CBC}, \feature{SHA1\_C}} & 1' & 1.00 \\
    & 14 &{\scriptsize \feature{CIPHER\_MODE\_STREAM}, \feature{AESNI\_C}, \feature{MD5\_C}} & 1' & 1.00 \\
    & 15 &{\scriptsize \feature{RIPEMD160\_C}, \feature{CIPHER\_MODE\_STREAM}} & 1' & 1.00 \\
    \midrule
    \multirow{3}{*}{\rotatebox[origin=c]{90}{\textbf{\tool{SQLite}}}} & 1 & {\scriptsize\feature{DEFAULT\_MEMSTATUS}, \feature{THREADSAFE}} & 1' & 1.00 \\
    & 2 & {\scriptsize\feature{DEFAULT\_MEMSTATUS}, \feature{MEMDEBUG},} & 1' & 1.00 \\
    & & {\scriptsize\feature{THREADSAFE}} & & \vspace{1ex}\\
    \bottomrule
  \end{tabular}
\end{table}

\subsubsection{Predicting Performance Interactions}
\label{sec:results-predicting}

\paragraph{\tool{mbedTLS} Direct Matching}
As we describe in Section~\ref{sec:fis}, one prediction method is to assume that every control-flow interaction induces a performance interaction that involves exactly the same features.
In \tool{mbedTLS}, from the 73 automatically identified unique control-flow interactions there are three---interactions 1, 3, and 4 in Table~\ref{tab:internal_in_external}---that have exactly the same features as the related performance interactions 2, 6, and 11 in Table~\ref{tab:result-perf-int}.
That is, three of the 16 performance interactions could be predicted by the direct matching.
Therefore, the precision of the direct matching is 4.11\,\% and the recall is 18.75\,\%. If we also incorporate the 11 indirect control-flow interactions, which we identified by manually inspecting the code, the total number of matching control-flow interactions becomes 14.
Including indirect control-flow interactions increases the precision and recall to 16.7\,\% and 51.85\,\% respectively.

\paragraph{\tool{SQLite} Direct Matching}
In \tool{SQLite}, there are no automatically identified unique control-flow interactions that match exactly any of the performance interactions. Including the manually added control-flow interactions gives the prediction precision of 5.13\,\% and the recall of 67\,\%.

\paragraph{\tool{mbedTLS} Frequent Item Sets}

Using frequent item set analysis (cf. Sec.~\ref{sec:fis}) on the set of control-flow interactions for \tool{mbedTLS}, we found 44 item sets, of which we calculated the support values.
The support values range from 11\,\% to 34\,\%, meaning that there are item sets occurring in 11\,\% to 34\,\% of all control-flow interactions.

Two of the found item sets match exactly the performance interactions 2 and 11 of Table~\ref{tab:result-perf-int}. Notice that we ran the frequent item set analysis only on the automatically detected control-flow interactions.
We were not able to run it on the indirect control-flow interactions, because then we would have to find every instance of such interaction manually, which is infeasible.
Nevertheless, we incorporated the indirect control-flow interactions into further analysis by approximating their support values based on the distribution of support values for similar indirect interactions (see Sec.~\ref{sec:threats}, for threats to validity). %
Among the 44 detected item sets, there are 33 item sets capturing interactions among ciphers, modes, and hash functions.
We assigned support values to the indirect control-flow interactions according to the distribution of the support values of these 33 item sets.
That is, 6\,\% of the interactions were assigned a support value of 11\,\%, 3\,\% were assigned a support value of 12\,\%, and so on.

By varying the threshold $E$, as described in Section~\ref{sec:fis}, we are able to decide which of the identified item sets are considered frequent.
By setting the threshold to 0, we consider all identified item sets as frequent.
When the threshold is increased the item sets with lower support values are not considered frequent anymore.
For example, if we set the threshold to 15\,\% only 25\,\% of the identified item sets will have a higher support value and will be considered frequent.
Changing the threshold this way allows us to observe its influence on the predictive power (i.e., precision and recall) of the frequent item sets.

To calculate how good the item sets are in predicting performance interactions, we compared how many of them denote the actually identified performance interactions (i.e., contain exactly the same features as the performance interactions).
The low precision and recall values for \tool{mbedTLS} summarized in Table~\ref{tab:precision-recall-fis} show that our predictor based on the frequent item sets has only a low predictive power.
Increasing the threshold value decreases the precision and recall of the predictor.

\paragraph{\tool{SQLite} Frequent Item Sets}
Applying the same frequent item set method to the control-flow interactions of \tool{SQLite} resulted in four frequent item sets with support values ranging from 20\,\% to 100\,\%. None of these frequent item sets matched the performance interactions. We could not approximate the distribution of the support values for the manually detected control-flow interactions, because they do not exhibit any commonalities with the calculated frequent item sets as it was the case for \tool{mbedTLS}.

\vspace{-1.5ex}\begin{framed}
\paragraph{Summary} We defined two predictors for performance interactions based on their relation with control-flow interactions.
The first predictor is based on the assumption that every control-flow interaction induces a performance interaction that involves exactly the same features.
The second predictor is based on the assumption that the recurring feature combinations in control-flow interactions capture the related performance interactions.
The evaluation showed that both predictors have only low precision and recall values.
\end{framed}

\subsection{Discussion}
\label{sec:discussion}

Based on our results, we conclude that there is indeed a quantifiable relation between control-flow and performance interactions.
We confirmed this by manually inspecting the code and by comparing which features are involved in the detected performance interactions and how these features interact at the control-flow level.
We found that features involved in performance interactions work closely together to implement the systems' functionality and also interact at the control-flow level.
That is, the same features that are involved in performance interactions are also involved in control-flow interactions.
Therefore, we can positively answer research question RQ1.\looseness=-1

The relation we found among control-flow and performance feature interactions has implications for performance prediction techniques for highly-configurable systems.
As we discussed in Section~\ref{sec:results-relating}, the identified control-flow interactions capture the features that are involved in the performance interactions.
Of course, we cannot identify these features precisely, because the same control-flow interactions also involve other features that are not involved in performance interactions (this is also a reason for direct matching prediction having low precision and recall; cf. Sec~\ref{sec:results-predicting}).
Nevertheless, assuming that only the features from the identified control-flow interactions can give rise to a performance interaction considerably reduces the search space of the potential performance feature interactions, because otherwise we have to assume that any (valid) feature combination may give rise to a performance interaction.
\tool{mbedTLS} has 134\,057 valid feature combinations of two and three features, but the 84 identified unique control-flow interactions (Sec.~\ref{sec:results-control-flow}) result in only 452 \emph{potential} performance interactions (among two and three features).
Notice that these include all 16 actually existing performance feature interactions that we identified.
That is, we are able to shrink the search space of performance feature interactions by almost 300 times (452 instead of 134\,057) without losing any of the actually existing performance feature interactions.
\tool{SQLite} has 524 valid feature combinations of two and three features and (based on the 39 identified unique control-flow interactions) only 131 \emph{potential} performance interactions (among two and three features). These potential performance interactions also include all 3 actually existing performance interactions that we identified. That is, the search space shrinks by 4 times.
These results have immediate consequences for performance prediction techniques based on machine learning and relying on sampling for building a training dataset: By exploiting our findings they can make sampling more focused on the configurations that potentially include interacting features, which may improve their prediction accuracy.

\begin{table}
  \centering
  \small
  \caption{Precision and recall values for the item sets as predictors for the performance interactions in \tool{mbedTLS}. (*)~marks the precision and recall values for the item sets with incorporated indirect control-flow interactions.\label{tab:precision-recall-fis}}
  \begin{tabular}{rrrrr}
    \toprule
    Threshold & Precision & Recall & Precision* & Recall*\\
    \midrule
    0  & 4.5 & 12.5 & 23.6 & 48.1 \\
    15 & 2.3 & 6.3  & 5.5  & 11.1 \\
    20 & 0   & 0    & 1.8  & 3.7 \\
    \bottomrule
  \end{tabular}
\end{table}

With respect to RQ2, we conclude that predictors based on direct matching and frequent item sets have low precision and recall values.
One possible reason is that the predictors rely solely on control-flow data, but features can also interact via data flow.
For example, they can exchange data through shared data structures.
This interplay at the data-flow level can be interpreted as data-flow feature interactions (much like control-flow feature interactions), which may also induce performance interactions.
For example, a feature may block other features by locking a shared data structure, which may have a negative influence on the performance of the system.
Therefore, enriching the data used by the predictors with the information about data-flow interactions may increase their predictive power.
So, a takeaway message here is that predictors should consider the interplay of features not only on the control-flow level, but also at the data-flow level, and other levels.
Another reason may be that not all features involved in a control-flow interaction are also involved in a related performance interaction.
The Jaccard index values in Table~\ref{tab:result-perf-int} show that only about half of the features in a control-flow interaction are also present in the related performance feature interaction.
For example, the interaction (\feature{AES\_C}, \feature{AESNI\_C}) has the average Jaccard index of 0.46.
This means that, on average, a related control-flow interaction has two other features additionally involved, in addition to features \feature{AES\_C} and \feature{AESNI\_C}.
Both predictors for a given control-flow interaction are not able to distinguish among features that are involved in a related performance interaction and those that are not.

\paragraph{Further Observations.}
A further observation is related to the distribution of the number of features involved in the control-flow and performance interactions.
For \tool{mbedTLS}, in most cases, interactions (both, control-flow and performance) involve two or three features.
For \tool{SQLite}, in most cases, control-flow interactions involve four features, but this is only the case because every single control-flow interaction involves the two crosscutting features  \feature{THREADSAFE} and \feature{ENABLE\_API\_ARMOR}.
If we ignore these crosscutting features, the pictures becomes similar to \tool{mbedTLS}.
The performance interactions in \tool{SQLite} involve two or three features as in \tool{mbedTLS}.
From these data, we conclude that the frequency of interactions decreases with the growing number of the involved features.
This shows that features tend to interact at the same rate (two or three features per interaction) independently of the type of the interaction (control-flow or performance).
This is another indication for a relation between control-flow and performance interactions.

Finally, for \tool{mbedTLS}, we found that most of the frequent item sets that we identified in the control-flow interactions contain features from three groups of algorithms: ciphers, modes, and hashes.
Even though most of the frequent item sets do not resemble existing performance interactions, they still capture the general pattern of the detected performance interactions, namely, that these interactions involve features from these three groups of algorithms.
For \tool{SQLite} the frequent item sets capture the crosscutting features, such as \feature{THREADSAFE} and \feature{ENABLE\_API\_ARMOR}. The crosscutting feature \feature{THREADSAFE} was involved in all identified performance interactions.

\section{Threats to Validity}
\label{sec:threats}
\paragraph{Internal Validity.}
Due to technical limitations of \tool{TypeChef}, we were unable to identify the exact number of indirect function calls between features (i.e., calls made using function pointers) and, consequently, the exact support values for the corresponding item sets (Sec.~\ref{sec:results-predicting}).
We approximated these support values based on the distribution of the support values for the item sets calculated from direct function calls.
Our approximation method may result in an inaccurate calculation of the precision and recall values of the frequent item set predictor.
Nevertheless, we expect that improving the approximation would rather improve the precision and recall of the predictor.

\paragraph{External Validity.}
Our study is exploratory in nature and aims at the initial investigation of the relation between control-flow and performance feature interactions.
Since we focused on analyzing two systems, our results may not hold for other highly-configurable systems.
Our study setup can serve as a blueprint for further studies that can rely on our conceptual framework for studying relations among external and internal interactions.

\section{Related Work}
In recent years, a number of papers aimed at detecting feature interactions in highly-configurable systems.
We summarize and subdivide them according to
our
classification~\cite{apel2013exploring}
into those considering internal feature interactions and those considering external feature interactions.
To our best knowledge, there is no work that study these two types of interactions in combination, as we do it in this study.

\paragraph{Internal Feature Interactions}
Detection of internal feature interactions is often used by techniques that aim at minimizing test-suite and test-effort for highly configurable systems.
Reisner et al.~\cite{reisner2010using}, Nguyen et al.~\cite{nguyen2016igen}, Tartler et al.~\cite{tartler2012configuration} apply symbolic evaluation, dynamic and static program analysis respectively to infer minimal sets of features responsible for a given code coverage. Kim et al.~\cite{kim2011reducing} apply static program analysis to identify features that do not interact with other features w.r.t. to the test-suite. Garvin et al. explores a connection between feature interactions and interaction faults~\cite{garvin2011feature}. Lillack et al. extends static taint analysis to automatically identify interactions among load-time configuration options~\cite{lillack2014tracking}.\looseness=-1

\paragraph{External Feature Interactions}
A number of recently proposed performance prediction techniques for highly configurable systems by Guo et al.~\cite{GC+13}, Siegmund et al.~\cite{siegmund2012predicting}, Sarkar et al.~\cite{sarkar2015cost}, Thereska et al.~\cite{thereska2010practical}, Westermann et al.~\cite{westermann2012automated}, and Zhang et al.~\cite{zhang2015performance} use machine-learning techniques, such as, CART, multivariate regression, and Fourier learning, for learning a performance function based on the performance measurements of a configuration sample.
These techniques learn performance (external) feature interactions as an integral part of the overall black-box learning process, that is, without considering the internal feature interactions.

\section{Conclusion}
In our case study we explored the relation among control-flow and performance feature interactions that occur in highly configurable systems.
Using the encryption library \tool{mbedTLS} and the database engine \tool{SQLite} as real-world subject systems, we identified control-flow and performance feature interactions using static program analysis and machine learning.
Analyzing the interactions, we found that they can be related based on the involved features.
By manually inspecting the code, we confirmed the causal relation between the interplay of features at the control-flow level and the identified performance interactions among the same features.
Furthermore, based on the identified relation, we defined two predictors for performance feature interactions and conducted a preliminary evaluation of these predictors.
The evaluation showed that the predictors have low precision and recall, presumably, because features also interact at the data-flow level.
Future predictors based on the internal feature interactions should consider both control-flow and data-flow interactions to improve their predictive power.

Beside this negative result, using the identified relation among control-flow and performance feature interactions, we are still able to shrink the search space of performance feature interactions (by almost 300 times for \tool{mbedTLS} and by 4 times for \tool{SQLite}) without losing any of the performance feature interactions actually existing in our subject systems.
Performance prediction techniques that rely on sampling can use our results to make their sampling more focused on configurations with potential performance interactions.

\section*{Acknowledgements}
  Kolesnikov’s, Grebhahn’s, and Apel’s work has been supported by the German Research Foundation (AP~206/5, AP~206/6, AP~206/7, AP~206/11) and by the Austrian Federal Ministry of Transport, Innovation and Technology (BMVIT) project No.~849928.
  Siegmund's work has been supported by the German Research Foundation under the contracts SI 2171/2 and SI 2171/3.
  Kästner’s work has been supported in part by the National Science Foundation (awards 1318808, 1552944, and 1717022), the Science of Security Lablet (H9823014C0140), and AFRL and DARPA (FA8750-16-2-0042).

\bibliographystyle{IEEEtran}
\bibliography{paper}

\end{document}